\documentclass[%
 reprint,
 amsmath,amssymb,
 aps,superscriptaddress
]{revtex4-1}

\usepackage{graphicx}
\usepackage{dcolumn}
\usepackage{bm}
\usepackage[version=3]{mhchem}
\usepackage[usenames, dvipsnames]{color}

\begin{document}

\preprint{APS/123-QED}

\title{Theoretical model for plasmonic photothermal response of gold nanostructures solutions}

\author{Anh D. Phan}
\affiliation{Department of Physics, University of Illinois, 1110 West Green St, Urbana, Illinois 61801, USA}
\affiliation{Institute of Physics, Vietnam Academy of Science and Technology, 10 Dao Tan, Hanoi 10000, Vietnam}
\email{adphan35@gmail.com}
\author{Do T. Nga}
\affiliation{Institute of Physics, Vietnam Academy of Science and Technology, 10 Dao Tan, Hanoi 10000, Vietnam}
\author{Nguyen A. Viet}
\affiliation{Institute of Physics, Vietnam Academy of Science and Technology, 10 Dao Tan, Hanoi 10000, Vietnam}

\date{\today}

\begin{abstract}
Photothermal effects of gold core-shell nanoparticles and nanorods dispersed in water are theoretically investigated using the transient bioheat equation and the extended Mie theory. Properly calculating the absorption cross section is an extremely crucial milestone to determine the elevation of solution temperature. The nanostructures are assumed to be randomly and uniformly distributed in the solution. Compared to previous experiments, our theoretical temperature increase during laser light illumination provides, in various systems, both reasonable qualitative and quantitative agreement. This approach can be a highly reliable tool to predict photothermal effects in experimentally unexplored structures. We also validate our approach and discuss its limitations.

\end{abstract}


\maketitle


\section{Introduction}
In recent years, much attention has been devoted to the usage of gold nanoshells and nanorods in photothermal therapy due to their wide range of applications
\cite{17,18,3,15}. Core-shell nanoparticles have been synthesized to merge the advantageous properties of these two materials and achieve plasmonic absorption
peaks in near-infrared regime. However, this design has been associated with particle size. Sub-100-nm nanoparticles are effective drug delivery carriers \cite{40,41,2}. Particles greater than 100 nm in diameter poorly penetrate membranes and may adhere to blood vessels and biological organs \cite{41}. While small gold nanorods have an additional absorption peak compared to their spherical counterparts \cite{1} within the near-infrared window, two surface plasmon resonances correspond to transverse and longitudinal modes. The desired position of absorbance maxima can be accurately tuned using current technological advancements \cite{7}. These gold nanostructures can absorb optical energy from incident laser light to heat and eradicate, locally and selectively, unhealthy cells with sizes less than 50 nm. The light-induced heating of gold nanoparticles generates acoustic waves which have been used in a new depth optoacoustic imaging technique with high resolution for non-invasive diagnostic techniques \cite{8,9}. The high efficiency of the surface plasmon absorption of gold nanostructures enhances the ultrasonic emission of the contrast. Gold nanoparticles can also penetrate cell membranes \cite{10,14} and are one of the most promising drug delivery vehicles.

\section{Theoretical Background}

In this work, we present a comprehensive model to describe the spatial and temporal temperature increase under laser exposure of gold nanoshells and nanorods in aqueous solutions. The absorbed optical energy is strongly dependent on the absorption cross section calculated using the extended Mie theory. 
\subsection{Mie theory for gold core-shell nanostructure}
Applying the Mie approach to the core-shell nanospheres provides \cite{22,23} 

\begin{eqnarray}
Q_{ext} &=& -\frac{2\pi}{k_m^2}\sum_{n=1}^{\infty}(2n+1)\ce{Re}\left(a_n + b_n \right), \nonumber\\
Q_{scat} &=& \frac{2\pi}{k_m^2}\sum_{n=1}^{\infty}(2n+1)\left(\left | a_n \right|^2  + \left | b_n \right |^2 \right),\nonumber\\
Q_{abs} &=& Q_{ext}  - Q_{scat},
\label{eq:1}
\end{eqnarray}
where 
\begin{eqnarray}
a_n&=&-\frac{U_n^{TM}}{U_n^{TM}+iV_n^{TM}} ,\quad b_n=-\frac{U_n^{TE}}{U_n^{TE}+iV_n^{TE}}, \nonumber\\
U_n^{TM} &=& \begin{vmatrix} j_n(k_cR_c) & j_n(k_sR_c) & y_n(k_sR_c) & 0 \\ \cfrac{\Psi_n^{'}(k_cR_c)}{\varepsilon_c} & \cfrac{\Psi_n^{'}(k_sR_c)}{\varepsilon_s} & \cfrac{\Phi_n^{'}(k_sR_c)}{\varepsilon_s}& 0 \\ 
0 & j_n(k_sR_s) & y_n(k_sR_s) & j_n(k_mR_s) \\
0 & \cfrac{\Psi_n^{'}(k_sR_s)}{\varepsilon_c} & \cfrac{\Phi_n^{'}(k_sR_s)}{\varepsilon_s} & \cfrac{\Psi_n^{'}(k_mR_s)}{\varepsilon_m}
\end{vmatrix}, \nonumber\\
V_n^{TM} &=& \begin{vmatrix} j_n(k_cR_c) & j_n(k_sR_c) & y_n(k_sR_c) & 0 \\ \cfrac{\Psi_n^{'}(k_cR_c)}{\varepsilon_c} & \cfrac{\Psi_n^{'}(k_sR_c)}{\varepsilon_s} & \cfrac{\Phi_n^{'}(k_sR_c)}{\varepsilon_s}& 0 \\ 
0 & j_n(k_sR_s) & y_n(k_sR_s) & y_n(k_mR_s) \\
0 & \cfrac{\Psi_n^{'}(k_sR_s)}{\varepsilon_c} & \cfrac{\Phi_n^{'}(k_sR_s)}{\varepsilon_s} & \cfrac{\Phi_n^{'}(k_mR_s)}{\varepsilon_m}
\end{vmatrix},\nonumber\\
\label{eq:2}
\end{eqnarray}
where $R_c$ and $R_s$ are the inner and outer radius of the core-shell
nanostructure, respectively, and $V_n$ and $U_n$ are determinants, $j_n(x)$ is the spherical Bessel function of the first kind, $y_n(x)$ is the spherical Neumann function, $\Psi(x)=xj_n(x)$ and $\xi_n(x) = xy_n(x)$ are the Riccati–Bessel functions. Analytical expressions of $U_n^{TE}$ and $V_n^{TE}$ are achieved by substituting the permeability for the dielectric function in Eq.(\ref{eq:2}). The dielectric functions of core, shell and surrounding (water) medium of the core-shell nanoparticles are $\varepsilon_c$, $\varepsilon_s$ and $\varepsilon_m \approx 1.77$, respectively. The wavenumber is $k_i=2\pi\sqrt{\varepsilon_i}/\lambda$ with $i=s, c$, and $m$. $\lambda$ is the wavelength of incident light in vacuum. In our calculations, the dielectric function of silica ($\varepsilon_c$) is taken from Ref.\cite{19}, while the dielectric function of gold ($\varepsilon_s$ or $\varepsilon_{Au}$) is described by the Lorentz-Drude model with several oscillators \cite{24}
\begin{eqnarray}
\varepsilon_{Au} = \varepsilon_s = 1- \frac{f\omega_p^2}{\omega^2-i\omega\Gamma_0} + \sum_{j=1}^{5}\frac{f_j\omega_p^2}{\omega_j^2-\omega^2+i\omega\Gamma_j},
\end{eqnarray}
where $f_0$ and $f_j$ are the oscillator strengths, $\omega_p$ is the plasma frequency for gold, and $\Gamma_0$ and $\Gamma_j$ are the damping parameters. All parameters in this model come from Ref. \cite{24}. However for the shell thickness less than 20 nm, the finite size effect becomes important. This effect can be added to the model by modifying the parameter $\Gamma_0 \equiv \Gamma_0+Bv_F/(R_s-R_c)$, where $v_F$ is the gold Fermi velocity, and $B$ is the parameter characterizing the scattering processes.

\subsection{Mie theory for gold nanospheroids}

The optical absorption cross section of a gold nanospheroid can be calculated by the extended Mie-Gans theory \cite{1}
\begin{eqnarray}
Q_{abs}=\frac{2\pi}{3\lambda}\sqrt{\varepsilon_m}Im(\alpha_{a}+\alpha_{b}+\alpha_{c}),
\label{1}
\end{eqnarray}
where $\alpha_j$ is the polarizability of the ellipsoid along $j$ direction,
where $j=a,b,c$ refers to characteristic lengths of the ellipse, and $a$ and $c$
are the semi-major and semi-minor axis, respectively. The calculations have also been widely applied to explain the absorption spectrum of nanorod as the spheroid is prolate. According to the extended Mie theory \cite{1}, $\alpha_j$ is expressed by

\begin{eqnarray}
\alpha_j(\omega) = \frac{4\pi abc}{3}\frac{\varepsilon_{Au}(\omega)-\varepsilon_m}{\varepsilon_m + L_j(\varepsilon_{Au}(\omega)-\varepsilon_m)},
\end{eqnarray}
where $L_j$ is a factor responsible for the ellipsoid shape.

For an prolate spheroid ($a > b = c$),
\begin{eqnarray}
L_a = \frac{1-e^2}{e^2}\left[\frac{1}{2e}\ln\left(\frac{1+e}{1-e} \right) -1\right],  L_{b,c} = \frac{1-L_a}{2},
\end{eqnarray}
where $e$ is the ellipticity given by $e=\sqrt[]{1-c^2/a^2}$.

For oblate spheroid ($a = b > c$),
\begin{eqnarray}
L_a &=& \frac{\sqrt{1-e^2}}{2e^3}\left[\frac{\pi}{2}-\tan^{-1}\left(\frac{\sqrt{1-e^2}}{e} \right)\right]-\frac{1-e^2}{2e^2},\nonumber\\
L_b&=& L_a,\quad L_c = 1-2L_a.
\end{eqnarray}

The damping parameter of gold nanorods is given by $\Gamma_0 \equiv \Gamma_0+Bv_F/L_{eff}$, where $L_{eff}=4V_s/S$, $V_s$ is the volume and $S$ is the surface area of the nanospheroids \cite{33}.

\subsection{Plasmonic heating}
Laser light with a wavelength in the NIR region can penetrate water and
biological tissue to excite the surface localized resonance of metal
nanostructures. In our calculations, we use the 808 nm laser light, similar to
most experiments. The nanostructures absorb optical energy and dissipate to the heat. We assume that the efficiency of the light-to-heat conversion is 100 $\%$ and the particles are randomly distributed in an effective spherical region of radius $R$. The radius $R$ is estimated by a radius of a sphere with a volume of measured suspensions. The temperature rise in the experimental samples under the laser light illumination is described by two distinct but strongly correlated processes: heat dissipation from particles and the convective heat transfer in medium. The time-dependent temperature increase $\Delta T(r,t)$ is theoretically modeled by the Pennes bioheat transfer equation \cite{25,26} in spherical coordinates
\begin{eqnarray}
\frac{1}{\kappa}\frac{\partial \Delta T}{\partial t} = \frac{1}{r^2}\frac{\partial}{\partial r}\left(r^2\frac{\partial \Delta T}{\partial r} \right) - \frac{\Delta T}{\kappa\tau} + \frac{A}{k},
\label{eq:3}
\end{eqnarray}
where $k$ and $\kappa=k/(\rho c)$ are the thermal conductivity and thermal
diffusivity of the medium, respectively,  $\rho$ is the mass density, $c$ is  the specific heat, $\tau$ is the perfusion time constant, $A = NQ_{abs}I_0$ is the heat source density due to absorbed energy on metal nanoparticles, $N$ is the number of particles per unit volume in the samples, and $I_0$ is an irradiation intensity. The heat source generation is thermally localized within the effective spherical region. Outside the area, there is no absorber to generate heat. Thus, Eq.(\ref{eq:3}) can be recast by
\begin{eqnarray}
\frac{1}{\kappa}\frac{\partial \Delta T}{\partial t} &=& \frac{1}{r^2}\frac{\partial}{\partial r}\left(r^2\frac{\partial \Delta T}{\partial r} \right) - \frac{\Delta T}{\kappa\tau} + \frac{A}{k}, \quad 0 \leq r \leq R, \nonumber\\
\frac{1}{\kappa}\frac{\partial \Delta T}{\partial t} &=& \frac{1}{r^2}\frac{\partial}{\partial r}\left(r^2\frac{\partial \Delta T}{\partial r} \right) - \frac{\Delta T}{\kappa\tau}, \qquad R \leq r.
\label{eq:4}
\end{eqnarray}

The temperature variation and its spatial derivatives have to be continuous at $r = R$. General solution of Eq.(\ref{eq:4}) can be analytically solved with the these boundary conditions to obtain temporal and spatial temperature distributions in the spherical region. The temperature at the center of the localized spherical domain of nanoparticles is assumed to be measured using thermal probes \cite{25,26}. The temperature at $r = 0$ is

\begin{eqnarray}
\Delta  T(r=0,t)&=&\frac{A}{k}\left[ -\kappa\int_0^t e^{-t'/\tau}\mbox{erfc}\left(\frac{R}{2\sqrt{\kappa t'}} \right)dt' \right. \nonumber\\
&-& \left. R\int_0^t e^{-t'/\tau}\sqrt{\frac{\kappa}{\pi t'}}\exp\left(-\frac{R^2}{4\kappa t'}\right)dt' \right.\nonumber\\
&+& \left. \kappa\tau(1-e^{-t/\tau}) \right].
\label{eq:5}
\end{eqnarray}

\section{Numerical results and discussions}
To verify the validity of our theoretical analysis, we compare our calculations with prior experiments \cite{27,28,29} which were carried out for gold nanoshells and nanorods dispersed in water. The perfusion time of water is approximately 2000 s. For silica-gold core-shell nanoparticles, the parameter $B$ is fixed at 1.5 while $B$ can be 0 or 0.33 for gold nanorods \cite{33}.

\begin{figure}[htp]
\includegraphics[width=8cm]{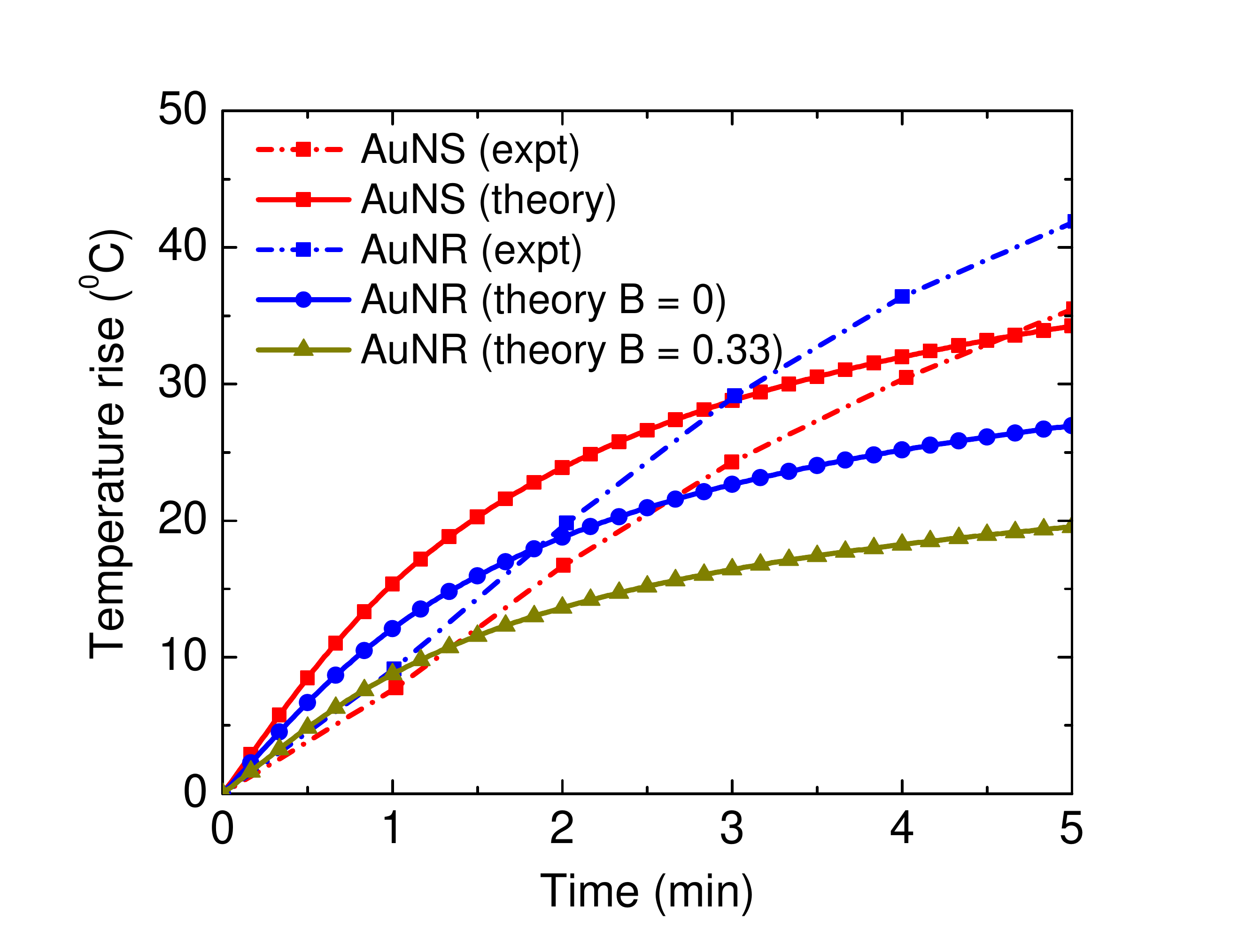}
\caption{\label{fig:1}(Color online) Temperature rise calculated by our theoretical model (solid lines) and experiments (dashed-dotted lines) for hybrid silica-gold core-shell nanoparticles with the core size of 120 nm and the gold shell thickness of 15 nm synthesized in Ref.\cite{27}. The concentration $N \approx 3.5\times10^{10}$ particles/ml and the volume $V = 2$ ml and $I_0 =3$ $W/cm^2$.}
\end{figure}

Figure \ref{fig:1} shows both theoretical calculations and measurements for the temperature rise in Ref.\cite{27}. At wavelength of 808 nm, $Q_{abs} \approx 42000$ $nm^2$ for the gold nanoshells, and the absorption cross sections of the gold nanorods with $B=0$ and 0.33 are 910 and 660 $nm^2$, respectively. Our numerical results of the core-shell composite agree relatively well with experiments. The quantitative deviation can originate from modeling the solution droplet containing the core-shell nanostructures in the experiment as a perfect sphere with radius $R$. However, the fact that the droplet is deposited on substrate and is not perfectly spherical.  Whereas there is qualitative but not quantitative agreement between the theory and experiments of the localized plasmonic heating of the prolate spheroids. The discrepancies are due to modeling the nanorod as a prolate spheroid and neglecting the particle size distribution in our calculations. The shape of spheroidal nanoparticles are less square than that of nanorods. Our numerical predictions of the plasmonic peak of gold nanorods at around 770 nm (close to experiments) is strongly sensitive to the nanostructure size and has much smaller bandwidth than experiments. Thus, using the Mie-Gans theory to compute the absorption of gold nanorods at 808 nm causes impreciseness in spite of the fact that the theory can plausibly predict the position and amplitude of optical maxima.   

\begin{figure}[htp]
\includegraphics[width=8cm]{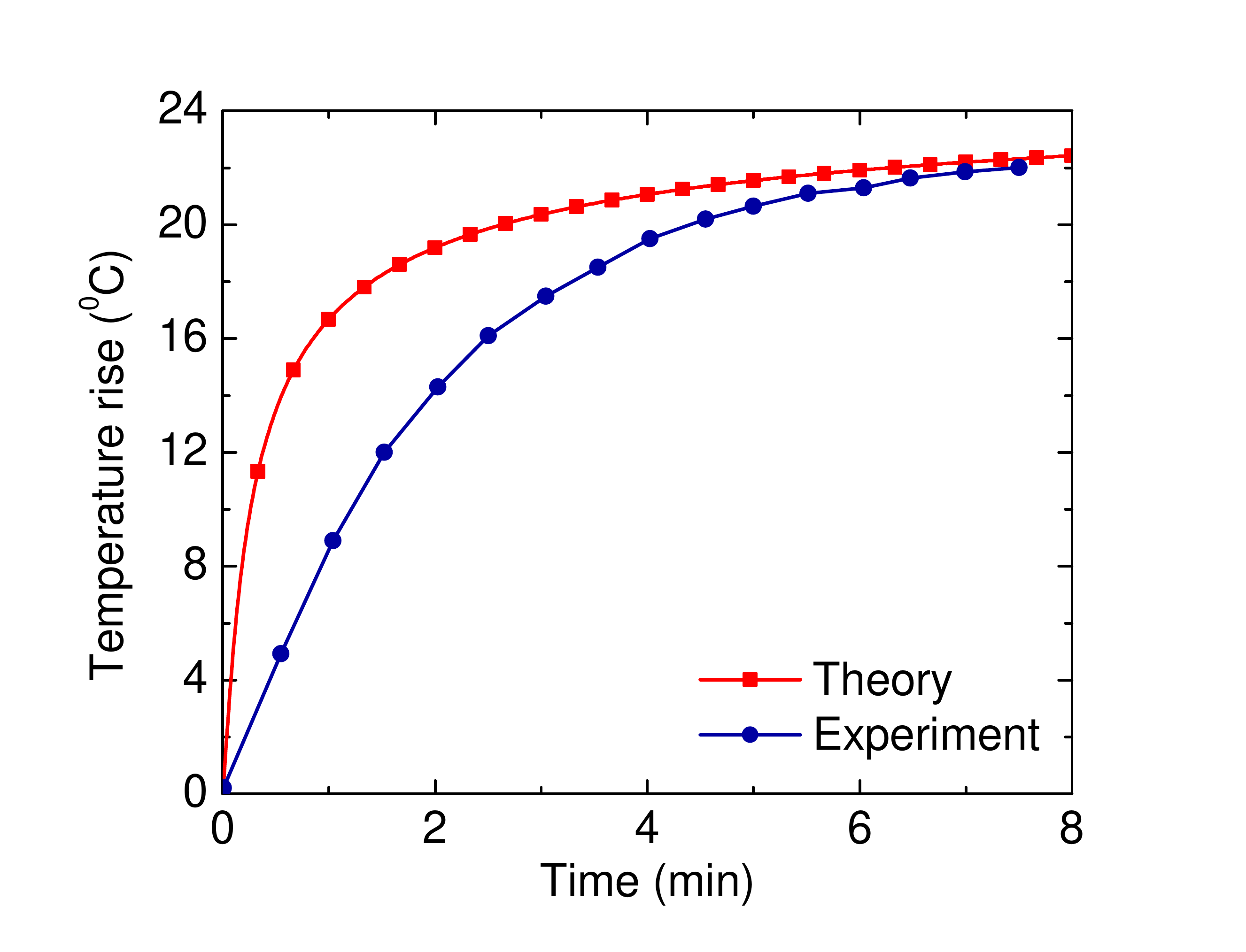}
\caption{\label{fig:2}(Color online) Temperature variation of aqueous suspensions of gold nanorods with an average diameter of 9.1 nm on the minor axis and a longitudinal size of roundly 36 nm prepared and characterized in Ref.\cite{28}. The concentration $N \approx 8.4322\times10^{11}$ particles/ml, the volume $V = 100$ $\mu$l and $I_0 =0.8$ $W/cm^2$.}
\end{figure}

To justify the validity of our approach to gold nanorod systems, we compare the theoretical analysis with different experiments in Ref.\cite{28} and present in
Fig.\ref{fig:2}. The computations are carried out by setting $B = 0.33$ to obtain the localized surface plasmon resonance at 808 nm and $Q_{abs}=553$
$nm^2$. The numerical results provide both qualitative and quantitative agreement experimental data. This finding suggests that our model for gold
nanorod system works well when the wavelength of the illuminating laser light coincides with the peak plasmon resonance.
\begin{figure}[htp]
\includegraphics[width=8cm]{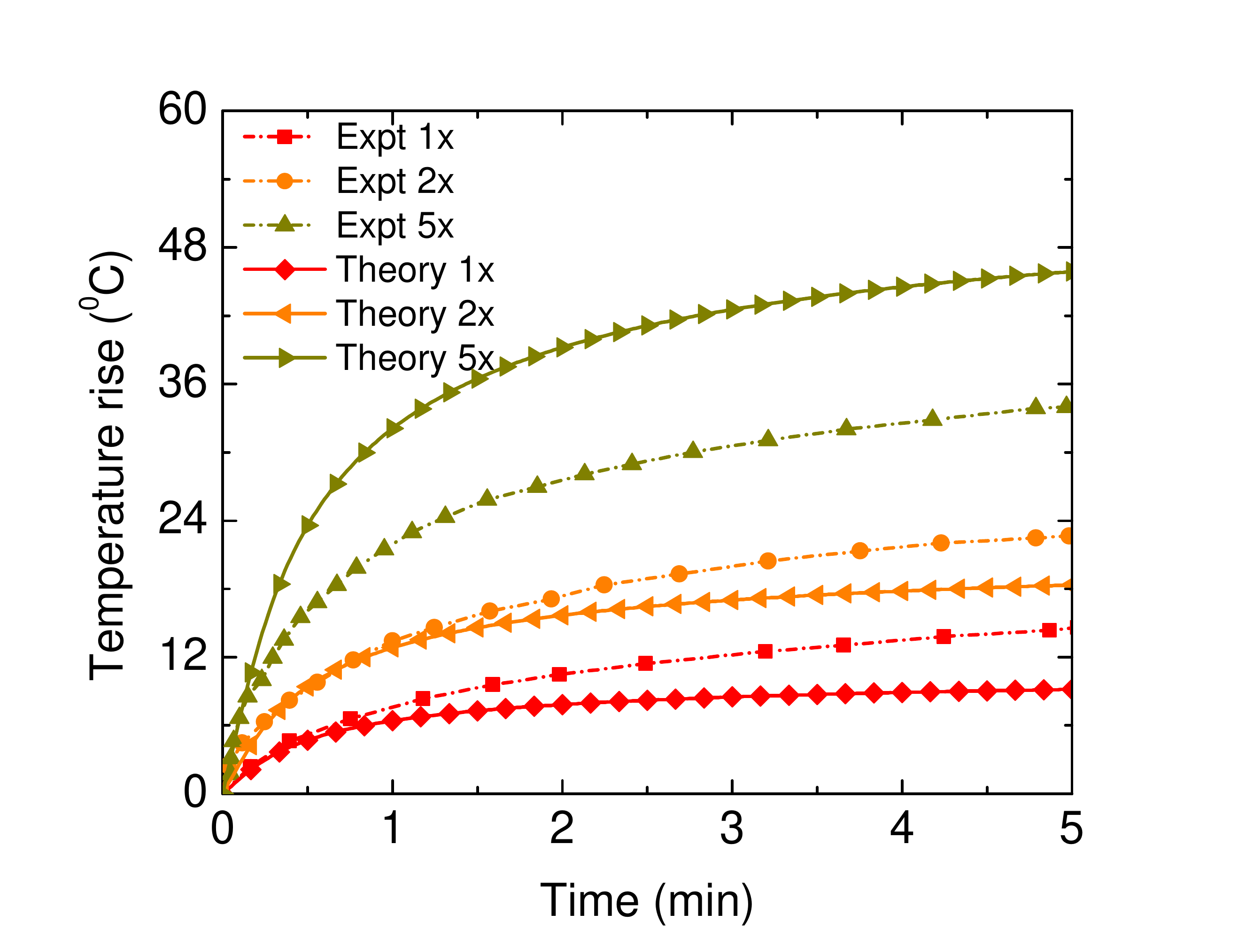}
\caption{\label{fig:3}(Color online) Temperature change calculated by our theoretical model (solid lines) and experiments (dashed-dotted lines) for nanoshells with 20 nm gold thickness on a silica core with a radius of 60 nm studied at various concentrations in Ref.\cite{29}. The concentration $x = 1.14\times10^{9}$ particles/ml, the volume $V = 260$ $\mu$l and $I_0 =2$ $W/cm^2$.}
\end{figure}

Different comparisons between our proposed theoretical model for gold nanoshell suspensions with experiments in Ref.\cite{29} implemented at different concentrations ($1x$, $2x$ and $5x$) are shown in Fig.\ref{fig:3}. The scattering Mie calculations predict $Q_{abs} = 40800$ $nm^2$ at 808 nm. Passable agreements between theory and experiments is readily apparent by observing the $1x$ and $2x$ concentration of the gold nanostructure. At higher densities, the disagreement becomes increasingly more remarkable. There are several explanations for the phenomemon. Our analysis reveals the linear increase of the steady-state temperature rise as the laser intensity is increased corresponding with conclusions found in previous theoretical and experimental studies \cite{26,30,31,32}. However, the experimental data in Fig.\ref{fig:3} contradicts this finding. One possibility is that experimental errors influence the reliability of Ref.\cite{29}. Another possibility is that our assumption of perfect light-to-heat conversion and disregard of the plasmonic scattering effects may be invalid in some cases. Authors in a prior work suggest that the nearly perfect energy conversion efficiency is expected for small nanoparticles, whose the scattering cross section is several orders of magnitude smaller as compared the absorption cross section \cite{1}. For larger nanoparticles, the scattering effects on the plasmonic heating become dominant and the transduction efficiency of energy conversion is expected to be smaller. These effects can be a direct consequence of the deviation between theory and experiment. Systematic future experiments are needed to confirm the limitation of our model. 

\section{Conclusions}
In summary, we have shown the theoretical analysis for optical properties and photothermal heating of gold nanorods and nanoshells dispersed in water. The absorption cross section of the nanostructures is investigated using the Mie and Gans theory. Gold nanoshells and nanorods effectively harvest optical energy of near-infrared light at 808 nm in different laser intensities and are assumed to perfectly dissipate the energy into solvent medium. Under laser light illumination, we predict an approximate linear steady-state temperature increase with the growth of the laser intensity and nanostructure concentration. Our numerical results agree well with experimental data of previous papers. In the case of gold nanorods, illuminating the system with a laser light of a non-resonant wavelength can trigger considerable difference between our approach and experiments. Experimental error is also a possible main reason for the discrepancy and it is necessary to have more experimental studies in the future to benchmark our photothermal model.

\begin{acknowledgments}
This work was supported by Vietnam National Foundation for Science and Technology Development (NAFOSTED) under grant number 103.01-2015.42. We would like to dedicate this work to the recently deceased Prof. Nguyen A. Viet who co-authored this paper.
\end{acknowledgments}

\end{document}